\documentclass[utf8]{frontiersSCNS} 

\usepackage{url,hyperref,microtype,subcaption}
\usepackage[onehalfspacing]{setspace}

\def\keyFont{\fontsize{8}{11}\helveticabold }
\def\firstAuthorLast{Berton, M. {et~al.}} 

\def\Address{$^{1}$ Dipartimento di Fisica e Astronomia "G. Galilei", Universit\`a di Padova, Vicolo dell'osservatorio 3, 35122, Padova (PD), Italy \\
$^{2}$ INAF - Osservatorio Astronomico di Brera, via E. Bianchi 46, 23807 Merate (LC), Italy  }
\def\corrAuthor{Corresponding Author}

\def\corrEmail{marco.berton@unipd.it}

\begin{document}
\onecolumn
\firstpage{1}

\title[Young AGN unification]{An orientation-based unification of young jetted AGN: the case of 3C 286} 

\author[\firstAuthorLast]{Berton, M.$^{1,2,*}$, Foschini, L.$^2$, Caccianiga, A.$^2$, Ciroi, S.$^1$, Congiu, E.$^{1,2}$, \\ Cracco, V.$^1$, Frezzato, M.$^1$, La Mura, G.$^1$, Rafanelli, P.$^1$} 
\address{} 
\correspondance{} 

\extraAuth{}

\maketitle

\begin{abstract}

\section{}
In recent years, the old paradigm according to which only high-mass black holes can launch powerful relativistic jets in active galactic nuclei (AGN) has begun to crumble. The discovery of $\gamma$-rays coming from narrow-line Seyfert 1 galaxies (NLS1s), usually considered young and growing AGN harboring a central black hole with mass typically lower than 10$^8$ M$_\odot$, indicated that also these low-mass AGN can produce powerful relativistic jets. The search for parent population of $\gamma$-ray emitting NLS1s revealed their connection with compact steep-spectrum sources (CSS). In this proceeding we present a review of the current knowledge of these sources, we present the new important case of 3C 286, classified here for the fist time as NLS1, and we finally provide a tentative orientation based unification of NLS1s and CSS sources.

\tiny
 \keyFont{ \section{Keywords:} Active Galactic Nuclei (AGN), narrow-line Seyfert 1 galaxies, compact steep-spectrum sources, relativistic jets, unification models} 
\end{abstract}

\section{Overview}
Relativistic jets are usually thought to be a product of accretion onto a central supermassive black hole (SMBH) \citep[e.g.][]{BlandfordZnajek, BlandfordPayne}. In active galactic nuclei (AGN), the interplay between the accretion disk and the SMBH is likely to produce powerful and collimated bipolar outflows. According to the unified model of jetted-AGN \citep{Urry95}, the two classes of blazars, BL Lacertae objects (BL Lacs) and flat-spectrum radio quasars (FSRQs), are double-lobed radio galaxies, with Fanaroff-Riley (FR) morphological types I and II, respectively, in which the line of sight falls inside the relativistic jet aperture cone. In recent years, this scenario has been slightly revised in a more physical way. BL Lacs and FSRQs indeed seem to reflect two different accretion modes. While BL Lacs have an inefficient accretion onto their central SMBH and a low density environment, FSRQs have instead a strong disk accreting efficiently and a photon- and matter-rich environment \citep{Best12}. This dichotomy led to the distinction between high-excitation radio galaxies and low-excitation radio galaxies (HERGs and LERGs, respectively) classified according, for example, to the [O III]/H$\alpha$ flux ratio \citep{Laing94}, and corresponding, respectively, to FSRQ and BL Lac objects when observed at small angles (although exceptions are known, \citealp{Giommi12}, see also the review by \citealp{Foschini17}). 

Another important aspect of jetted-AGN is that for many years they were thought to be produced only by the most massive black holes harbored in giant passive elliptical galaxies \citep{Laor00, Boettcher02, Marscher09}. However, the discovery of gamma-ray emission coming from the AGN class known as narrow-line Seyfert 1 galaxies (NLS1s) proved this paradigm to be wrong \citep{Abdo09a, Abdo09b, Foschini15}. Low mass AGN, indeed, are also able to launch relativistic jets, whose power is lower because of the well-known scaling of the black hole mass \citep{Heinz03, Foschini11b}. This result was confirmed by the identification of low-mass sources among flat-spectrum radio quasars (FSRQs, \citealp{Ghisellini16}), and by the discovery of relativistic jets associated with spiral galaxies \citep[e.g.][]{Keel06, Mao15}. \par

NLS1s are the most prominent class of low-mass AGN. Classified according to the FWHM(H$\beta$) $<$ 2000 km s$^{-1}$ and the ratio between [O III]/H$\beta <$ 3, they are unobscured AGN with relatively low black hole mass ($\lesssim$ 10$^8$ M$_\odot$) and high Eddington ratio, which accounts for the narrowness of the permitted lines and other observational properties \citep[for a review, see][]{Komossa08a}. A well-known interpretation for these peculiar features is that NLS1s are young AGN still growing and evolving \citep{Mathur00}, or sources rejuvenated by a recent merger \citep{Mathur12}. While a large fraction of NLS1s is radio-quiet, some of them are radio-loud and, as previously mentioned, harbor powerful relativistic jets \citep{Yuan08}. These radio-loud NLS1s (RLNLS1s) are characterized by a strong and very fast variability consistent with the low-mass scenario, and particularly evident in those with a flat radio spectrum \citep{Foschini15}. 

Another class of radio-loud AGN is known as compact steep-spectrum sources (CSS, \citealp[see the review by][]{Odea98}). The radio morphology of these objects is typically characterized by fully developed radio lobes and by a small linear size, with jets confined within the host galaxy. Many authors interpret these properties as signs of young age \citep{Fanti95}: different estimates of this parameter indeed revealed that they could be radio sources younger than 10$^5$ yrs \citep{Owsianik98, Murgia99}. CSS sources might eventually evolve to giant double-lobed radio galaxies, or simply switch off because of disk instabilities, going through multiple activity phases during their lifetime \citep[e.g.][]{Czerny09, Orienti15}. As regular radio galaxies, CSS sources can also be divided in HERGs and LERGs according to their optical spectrum \citep{Kunert10a, Kunert10b}.

\section{CSS sources and NLS1s}
A possible link between NLS1s and CSS was first suggested by \citet{Oshlack01}, who remarked how the RLNLS1 PKS 2004-447 radio morphology was consistent with that of a CSS source. In the following years, many authors reached the same conclusion \citep{Komossa06, Gallo06a, Yuan08, Caccianiga14, Schulz15}. In particular \citet{Gu15} carried out a VLBA survey on 14 NLS1s at 5 GHz, finding that essentially all RLNLS1s with a steep radio spectrum have a compact morphology, just like CSS. \citet{Caccianiga17}, focusing on the steep-spectrum RLNLS1 SDSS J143244.91+301435.3, found a spectral turnover at low frequency and a small linear size, confirming that this object can be classified as CSS as well.  \par

Beside the morphological similarities, there are other hints that point toward a unification between CSS sources and RLNLS1s. This was investigated under three different points of view in \citet{Berton16c}, using two complete samples of CSS/HERGs and flat-spectrum RLNLS1s (F-NLS1s) limited to z $=$ 0.6. A first indication of similarity between these two samples was provided by the V/V$_{max}$ test, which revealed that both of them are consistent with no evolution up to their redshift limit. Moreover, the 1.4 GHz luminosity function (LF) of both samples was derived. Following \citet{Urry95}, the effect of relativistic beaming was applied to the observed LF of CSS/HERG to predict the expected shape of the LF of the sources observed along the jet direction. The resulting LF was in good agreement with the observed LF of F-NLS1 thus indicating that at least some CSS/HERGs might be RLNLS1s observed at large angles. 

Finally, a comparison between the black hole mass distributions of both samples unveiled again a very good agreement. We repeated here the analysis on the black hole mass distributions on a larger sample of 60 NLS1s. In Fig.~\ref{fig:2} we show the distribution of the CSS/HERGs sample by \citet{Berton16c}, compared to all of the masses derived for flat- and steep-spectrum RLNLS1s in \citet{Foschini15} (42 objects) and \citet{Berton15a} (17 sources), and to that of 3C 286 (see Sect.~\ref{sec:3C286}). It is immediately evident from the plot that 3C 286 is located in the same region as NLS1s. To test the similarity of the distributions, we applied the Kolmogorov-Smirnov test (K-S). The null hypothesis is that the two distributions are originated by the same population. As a rejection threshold, we used a p-value lower than 0.10. Considering CSS/HERGs, the null hypothesis cannot be rejected when comparing them to both flat- and steep-spectrum RLNLS1s, with p-values 0.82 and 0.39, respectively. Testing RLNLS1s as a whole, the p-value is 0.68, again too high to reject the null hypothesis. The mass distribution of CSS/HERGs is therefore statistically consistent with that of RLNLS1s, in agreement with the previously shown results. 

\section{3C 286}
\label{sec:3C286}
We present here another interesting case of CSS/HERG, that of 3C 286 (z = 0.85). This object is a well known CSS classified for the first time by \citet{Peacock82}. It is often used as a calibrator at centimeter wavelengths for both total flux and linear polarization. It shows both the turnover in the radio spectrum around 300 MHz, a spectral index between 1.4 and 50 GHz of $\alpha = -0.61$. It also exhibits the compact morphology typical of CSS: while it remains unresolved at kpc scales, a core-jet structure is visible on pc scale. The jet inclination has been estimated of 48$^\circ$ with respect to the line of sight \citep[e.g., see][and references therein]{An17}. 

\begin{figure}[h!]
\begin{center}
\includegraphics[width=10cm]{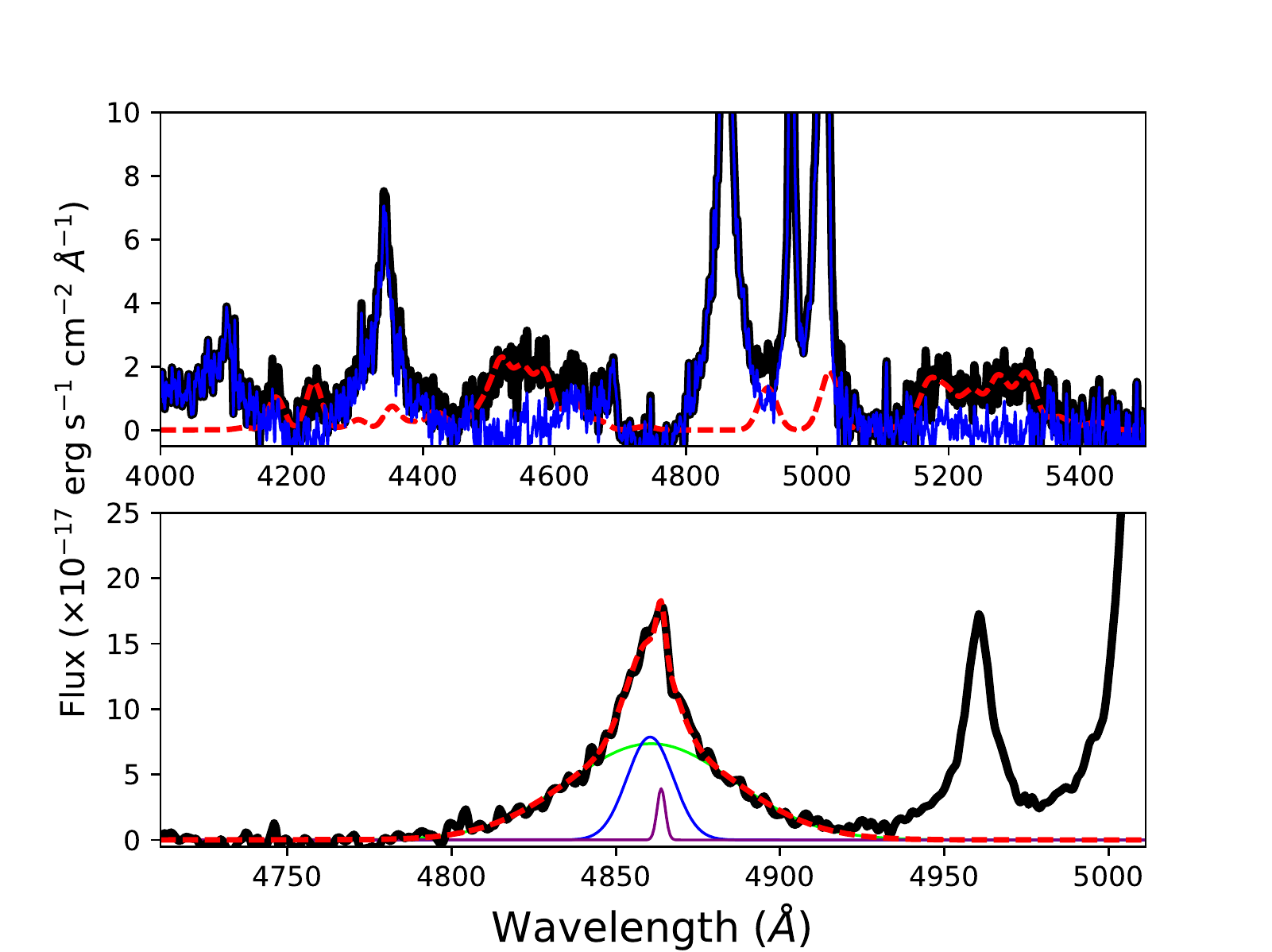}
\end{center}
\caption{Section of the optical spectrum of 3C 286, obtained from the SDSS-BOSS survey. \textbf{Top panel:} black solid line is the region between 4000 and 5500 \AA{} restframe continuum subtracted, the red dashed line represents the Fe II template, while the blue solid line is the spectrum after the iron subtraction. \textbf{Bottom panel:} black solid line indicates the H$\beta$ region, the three Gaussians used to reproduce the line profile are indicated by the green, blue and purple solid line, and the red dashed line represents their sum.}
\label{fig:1}
\end{figure}

The first optical spectrum of this radio source was obtained by \citet{Schmidt62}, who did not recognize the strongly redshifted Mg II $\lambda$2798 line. Since its redshift determination \citep{Burbidge69}, a significant improvement has been made in optical with the SDSS-BOSS survey \citep{Dawson13}, which obtained a new spectrum extended up to 10000\AA{}. The H$\beta$ line, now visible, has a FWHM of (1811$\pm$169) km s$^{-1}$, a ratio [O III]/H$\beta$ of 1.03$\pm$0.05, and Fe II multiplets are clearly visible in the spectrum (see Fig.~\ref{fig:1}). The corresponding uncertainties were calculated with a Monte Carlo method described by \citep{Berton16b}. These parameters clearly indicate that 3C 286 can be classified as NLS1. \par

\begin{figure}[h!]
\begin{center}
\includegraphics[width=10cm]{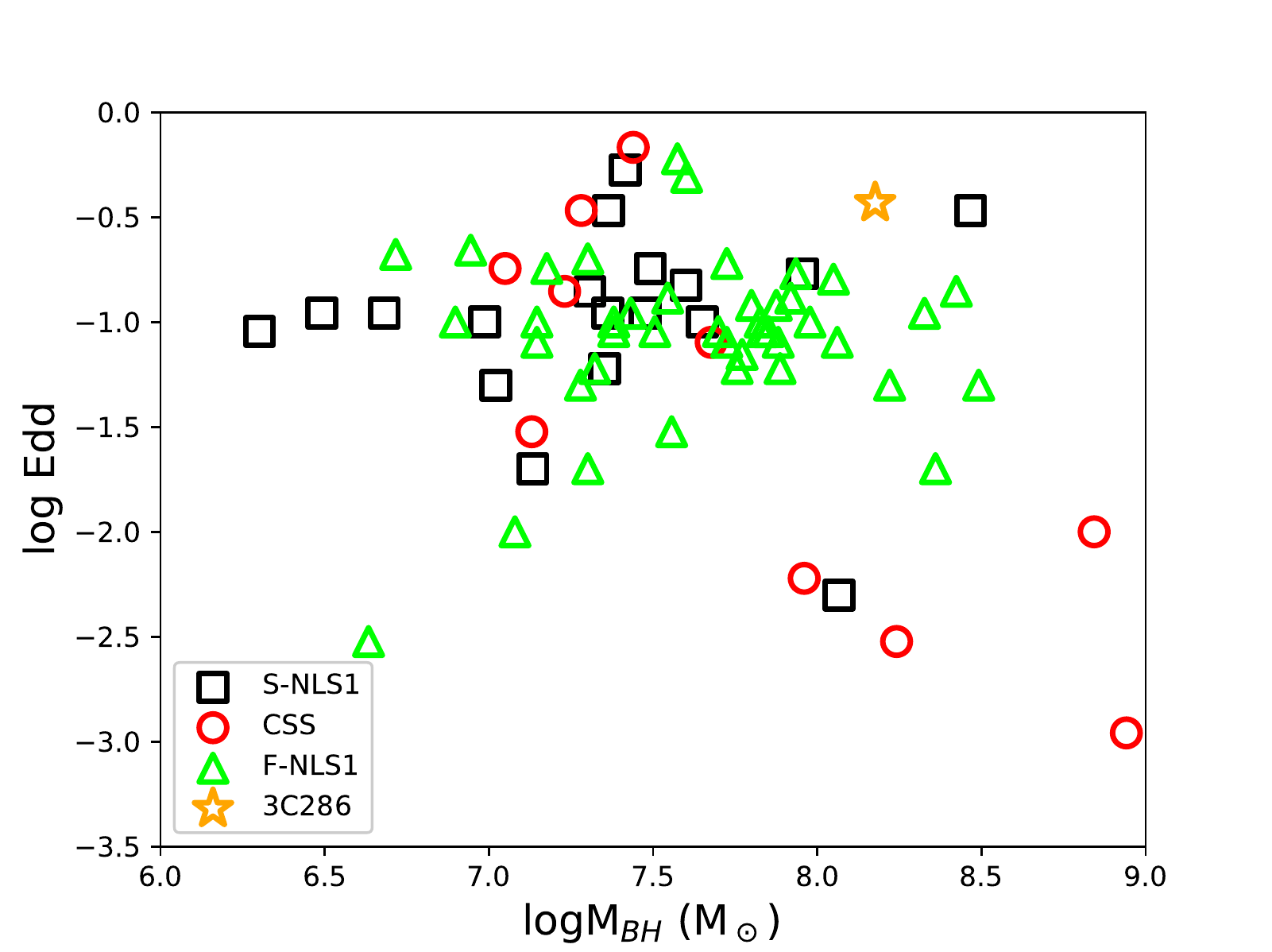}
\end{center}
\caption{Black hole mass vs Eddington ratio distribution. Flat-spectrum RLNLS1s are indicated by green triangles, the red circles indicate CSS/HERG sources, the black squares indicate steep-spectrum RLNLS1s, and the orange star represents 3C 286.}
\label{fig:2}
\end{figure}

We derived its black hole mass following the method described by \citet{Foschini15}. After correcting for redshift, we subtracted the continuum emission by fitting it with a power law, plus the Fe II multiplets using an online software \citep{Kovacevic10, Shapovalova12}\footnote{http://servo.aob.rs/FeII\_AGN/}, as shown in top panel of Fig.~\ref{fig:1}. We then fitted H$\beta$ profile using three Gaussians, one to reproduce the narrow component and two to reproduce the broad component (see bottom panel of Fig.~\ref{fig:2}). The narrow component flux was fixed to be 1/10 of [O III] $\lambda$5007 \citep{Veron01} and its FWHM to be the same as [O II] $\lambda$3727 ($\sim$600 km s$^{-1}$). After subtracting the H$\beta$ narrow component, we derived the second-order moment of the broad component $\sigma$ as a proxy for velocity \citep{Peterson04}, we obtained the broad-line region (BLR) radius by means of H$\beta$ luminosity. Under the hypothesis of virialized system, the black hole mass is 1.5$\times$10$^8$ M$_\odot$ and the Eddington ratio, defined as the ratio between the bolometric luminosity and the Eddington luminosity of the black hole, is 0.37. Both of these values are consistent with those derived for other radio-loud NLS1s (\citealp{Foschini15}, see Fig.~\ref{fig:2}). 

Interestingly, this object is a $\gamma$-ray emitter, identified in the 3rd Fermi catalog with the source 3FGL J1330.5+3023 \citep{Ackermann15}\footnote{http://www.asdc.asi.it/fermi3fgl/}. 3C 286 is therefore the third $\gamma$-ray detected steep-spectrum RLNLS1 after RX J2314.9+2243 \citep{Komossa15} and B3 1441+476 \citep{Liao15}. This is a very interesting result, since the number of misaligned NLS1s detected in $\gamma$-rays appears to be significantly higher than in other AGN classes. Indeed, only $\sim$10 radiogalaxies are known to be $\gamma$-ray emitters \citep{Rieger16}, and their corresponding beamed population of FSRQs and BL Lacs is made of 1144 sources. Conversely, among NLS1s, currently three misaligned $\gamma$-ray sources are known, out of nine oriented objects. The physical interpretation of this phenomenon is not clear yet, but interaction between the jet and the interstellar medium inside the host galaxy could in principle produce the observed $\gamma$-ray emission \citep{Migliori14}.

The classification of this source has another important implication on NLS1s nature. Some authors \citep[e.g.][]{Decarli08} suggested that NLS1s are not objects with a low-mass black hole and a high accretion rate. If their BLR had a disk-like shape, when observed pole-on the permitted lines would appear as narrow because of the lack of Doppler broadening. NLS1 would be just low-inclination sources rather AGN with a low-mass central SMBH. The large observing angle (48$^\circ$) estimated by \citet{An17} for 3C 286 seems to contradict this hypothesis, at least for this particular object and other RLNLS1s sharing the same property (e.g. Mrk 783, \citealp{Congiu17}).

\section{Young jetted-AGN unification}
As previously mentioned, CSS sources are usually considered as young radiogalaxies still growing \citep{Fanti95}. The previous results seem to point out that these objects can often be jetted-NLS1s observed at large angles. Of course, this does not mean that all CSS sources are misaligned NLS1s, because some objects probably have a larger black hole mass. The opposite instead might be true, since many of the jetted-NLS1 observed at large angles studied so far appear as CSS/HERGs. An important conclusion can be inferred from this result. If CSS/HERGs are young sources, and RLNLS1s are CSS sources, they should be young objects as well. This result, just like the high inclination of 3C 286, is in contrast with the vision of NLS1s as a pole-on view of a type 1 AGN. 

\begin{figure}[h!]
\begin{center}
\includegraphics[width=10cm]{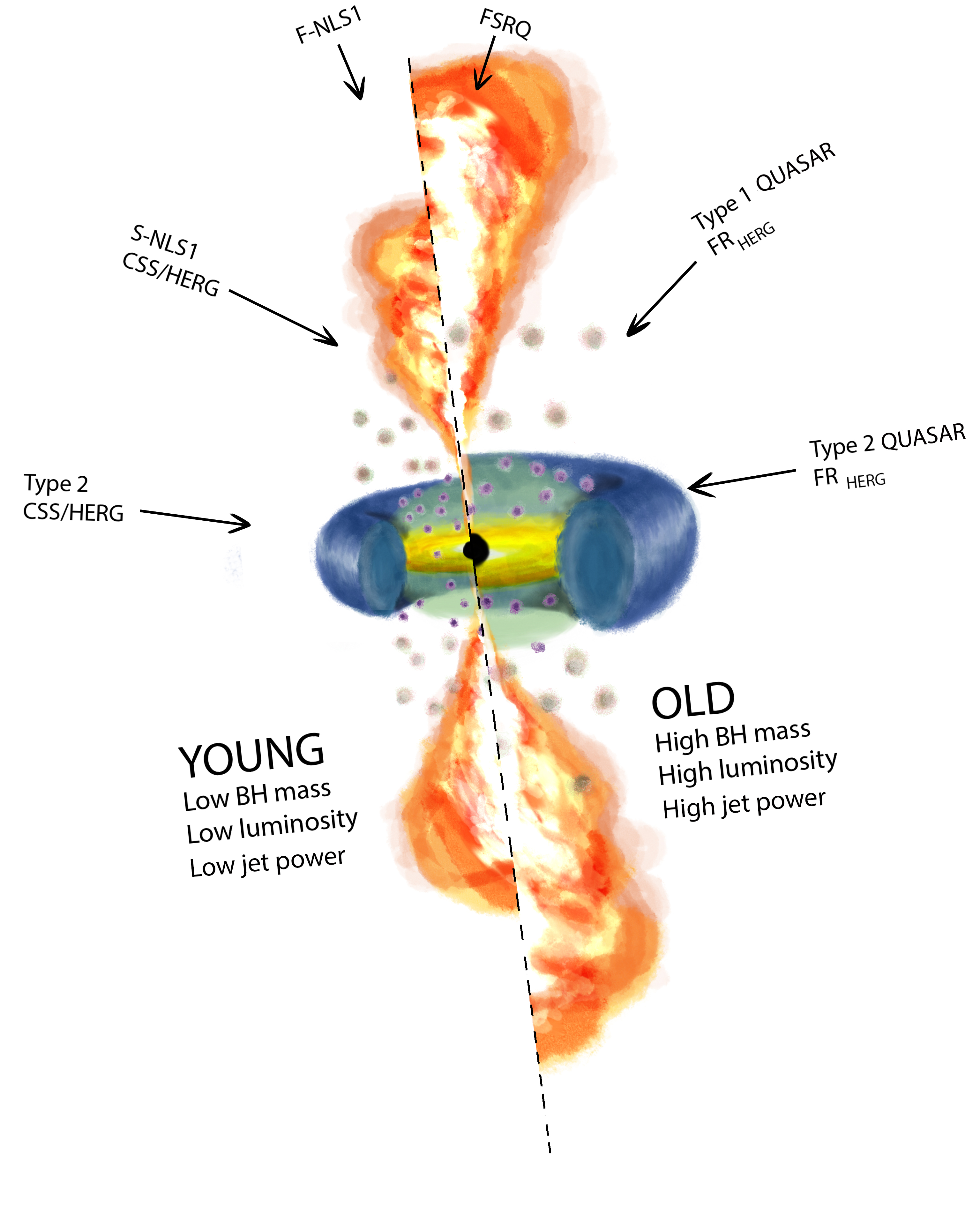}
\end{center}
\caption{Unification scheme of jetted-AGN with high accretion rate with respect to the Eddington limit and a high-density photon-rich environment. On the left side, young and smaller sources (NLS1s and CSS sources), compared to older and larger objects (FSRQs and FR$_{HERG}$).}
\label{fig:3}
\end{figure}

According to the unified model of radio-loud AGN \citep{Urry95}, the parent population of FSRQs are FR II radio galaxies, that is when an FR II is observed along its relativistic jet, it appears as a FSRQ. In a similar way, FR I radio galaxies constitute the parent population of BL Lacs. This interpretation might not be completely appropriate, since there are FR I sources that can be associated with FSRQs, and FR II which appear as BL Lacs. Recently it was suggested that a more accurate unification is between FSRQs and all the FR radio galaxies optically classified as HERG (FR$_{HERG}$), while BL Lacs should be associated to FR with a LERG-type optical spectrum (FR$_{LERG}$, \citealp{Giommi12}). This revised unification is more physics-based, since it accounts for the different accretion mechanisms (strong disk for FSRQs, weak disk for BL Lacs, \citealp{Best12}). CSS/HERGs, in this picture, are the young stage of FR$_{HERG}$ objects. Therefore, if flat-spectrum RLNLS1s are beamed CSS/HERGs, and FSRQs are beamed FR$_{HERG}$, F-NLS1s could be the young stage of FSRQs (\citealp{Foschini15, Berton16c}, see Fig.~\ref{fig:3}.)

This is in agreeement with the finding that flat-spectrum RLNLS1s are the low-mass tail of FSRQs \citep{Foschini15, Paliya16}. Indeed, the jet power scales with the black hole mass \citep{Heinz03, Foschini11b} and, as expected, the jet power of RLNLS1s is lower than that observed in FSRQs. Once rescaled for the central mass, though, the normalized power is essentially the same, indicating that the launching mechanism is the same \citep{Foschini15}. Finally, as shown in \citet{Berton16c}, the radio luminosity function of flat-spectrum RLNLS1s suggests that they might be the low-luminosity tail of FSRQs, indicating again an evolutionary link between these two classes. 

All these results can be summarized in the scheme of Fig~\ref{fig:3}. A young jetted-AGN with a strong accretion disk and a photon-rich environment in the central engine surroundings, when observed along its relativistic jets, can appear as a flat-spectrum RLNLS1, characterized by a low black hole mass and a low jet luminosity and power. When the inclination angle increases, the object appears first as a steep-spectrum RLNLS1, with an extended radio emission and a relatively small linear size, that can be classified as a CSS/HERG. Finally, when the line of sight intercepts the molecular torus surrounding the nucleus, the object appears as a type 2 AGN in optical and as a CSS in radio. For older jetted-AGN, instead, the usual unified model holds, with FR$_{HERG}$ (type 1 or 2 quasars in optical, according to the inclination) as parent population of FSRQs. 

Such a scheme is probably true only in a statistical sense. Objects like 3C 286 are still fairly rare, and might represent an exception. Moreover, it is unlikely that all type 1 (unobscured) young AGN fit the NLS1 classification. Several other non-NLS1 unobscured AGN with a low-mass black hole exist and were found among $\gamma$-ray emitting sources \citep[e.g.][]{Shaw12, Ghisellini16}. These objects do not appear as NLS1s even when observed at larger angles and in fact, as expected, not all type 1 CSS/HERGs can be classified as NLS1s. However, despite these notable exceptions, the scheme of Fig.~\ref{fig:3} seems to be consistent with several observations. 

This orientation-based unification could be confirmed by means of RLNLS1s and CSS/HERGs host galaxy observations. NLS1s usually have a late-type host galaxy \citep{Crenshaw03}, and some RLNLS1s indeed seem to share the same property \citep{Anton08, Caccianiga15, Kotilainen16, Olguiniglesias17}. CSS sources in general are believed to be instead hosted by early-type hosts \citep[e.g.][]{Best05}. However, some studies pointed out that their host galaxy can be a spiral as well \citep{Morganti11}, and that HERGs and LERGs can live in different morphological types, with HERGs showing more late-type hosts \citep{Best12}. Further observations are needed to better investigate this crucial aspect of young AGN. 

\section{Summary}

In this proceeding we presented a review of the current knowledge about young jetted-AGN, showing as well new results strictly connected to this topic and a tentative orientation-based unification model of these sources. We provided a new classification for the well-known CSS source 3C 286 which, thanks to new observations in the near-infrared, can be now classified as a NLS1 galaxy. This result confirms that RLNLS1s can often appear, in radio, as CSS sources. Moreover, we confirmed that the black hole mass distribution of CSS/HERGs is very similar to that of RLNLS1s. Along with other results already published in the literature, these findings seem to strengthen the scheme according to which flat-spectrum RLNLS1s are CSS/HERGs observed along their relativistic jet, that can appear as a steep-spectrum RLNLS1 when observed at intermediate angles, and as a type 2 radio galaxy when obscured. RLNLS1s and CSS/HERGs then might represent the young and growing stage of FSRQs and FR$_{HERG}$ radio galaxies, respectively, and constitute a part of the young AGN population. New observations on larger samples, particularly aimed at determining the host galaxy morphology, are needed to provide a final confirmation to this unification model.

\section*{Fundings}
This work has been partially supported by PRIN INAF 2014 \textit{Jet and astro-particle physics of gamma-ray blazars} (PI F. Tavecchio). 

\section*{Author Contributions}
The Author Contributions section is mandatory for all articles, including articles by sole authors. If an appropriate statement is not provided on submission, a standard one will be inserted during the production process. The Author Contributions statement must describe the contributions of individual authors referred to by their initials and, in doing so, all authors agree to be accountable for the content of the work. Please see  \href{http://home.frontiersin.org/about/author-guidelines#AuthorandContributors}{here} for full authorship criteria.

\section*{Acknowledgments}
We are grateful to Mrs. Lucia Zarantonello for drawing the sketch of the young unification scheme. This research has made use of the NASA/IPAC Extragalactic Database (NED) which is operated by the Jet Propulsion Laboratory, California Institute of Technology, under contract with the National Aeronautics and  Space Administration. Funding for the Sloan Digital Sky Survey has been provided by the Alfred P. Sloan Foundation, and the U.S. Department of Energy Office of Science. The SDSS web site is \texttt{http://www.sdss.org}. SDSS-III is managed by the Astrophysical Research Consortium for the Participating Institutions of the SDSS-III Collaboration including the University of Arizona, the Brazilian Participation Group, Brookhaven National Laboratory, Carnegie Mellon University, University of Florida, the French Participation Group, the German Participation Group, Harvard University, the Instituto de Astrofisica de Canarias, the Michigan State/Notre Dame/JINA Participation Group, Johns Hopkins University, Lawrence Berkeley National Laboratory, Max Planck Institute for Astrophysics, Max Planck Institute for Extraterrestrial Physics, New Mexico State University, University of Portsmouth, Princeton University, the Spanish Participation Group, University of Tokyo, University of Utah, Vanderbilt University, University of Virginia, University of Washington, and Yale University.

\bibliographystyle{frontiersinSCNS_ENG_HUMS} 
\bibliography{test}
\clearpage



\end{document}